\documentclass[twocolumn,times,twocolappendix,tighten,11pt,a4paper]{aastex631}
\usepackage{amsmath}
\usepackage{kantlipsum}
\usepackage{enumitem}
\usepackage{upgreek}
\usepackage{nicefrac}

\newcommand{\nablap}{\nabla_{\kern-.15em p}}

\shorttitle{}
\shortauthors{}

\graphicspath{{./}{figures/}}

\begin{document}

\title{\vspace*{-.4in}\Large Temperature Structures Associated with Different Components of the Atmospheric Circulation on Tidally Locked Exoplanets\vspace*{.1in}}

\author{Neil T. Lewis}
\author{Mark Hammond}
\affiliation{Atmospheric, Oceanic and Planetary Physics, University of Oxford, Oxford, UK}


\correspondingauthor{N.~T. Lewis}
\email{neil.lewis@physics.ox.ac.uk}

\begin{abstract}
Observations of time-resolved thermal emission from tidally locked exoplanets can tell us about their atmospheric temperature structure. Telescopes such as JWST and ARIEL will improve the quality and availability of these measurements. This motivates an improved understanding of the processes that determine atmospheric temperature structure, particularly atmospheric circulation. The circulation is important in determining atmospheric temperatures, not only through its ability to transport heat, but also because any circulation pattern needs to be balanced by horizontal pressure contrasts, therefore implying a particular temperature structure. In this work, we show how the global temperature field on a tidally locked planet can be decomposed into contributions that are balanced by different components of the atmospheric circulation. These are the superrotating jet, stationary Rossby waves, and the divergent circulation. To achieve this, we partition the geopotential field into components balanced by the divergent circulation and the rotational circulation, with the latter comprising the jet and Rossby waves. The partitioned geopotential then implies a corresponding partitioning of the temperature via the hydrostatic relation. We apply these diagnostics to idealised general circulation model simulations, to show how the separate rotational and divergent circulations together make up the total three-dimensional atmospheric temperature structure. We also show how each component contributes distinct signatures to the thermal phase curve of a tidally locked planet. We conclude that this decomposition is a physically meaningful separation of the temperature field that explains its global structure, and can be used to fit observations of thermal emission.
    \vspace*{-.35in}\\
\end{abstract}

\keywords{Exoplanets (498); Exoplanet atmospheres (487); Atmospheric Circulation (112) \vspace*{.2in}}

\section{Introduction} \label{sec:intro}

Recently, the collection of known planets has grown significantly. This is due to the discovery of numerous planets orbiting stars other than the Sun, called extrasolar planets (or \emph{exoplanets} for short).

The most observationally accessible exoplanets tend to be on close-in, short-period orbits. Many are sufficiently close to their host star that the gravitational interaction between star and planet synchronises their orbital and rotational periods \citep{dole1964locking,guillot1996locking}. A planet in this orbital state is referred to as \emph{tidally locked}. As tidally locked planets always present the same face to their host star, they have a permanent `day side' and `night side'.

These planets can be characterised by observing their thermal emission \citep{2014Natur.513..345B,2015PASP..127..941C}. Time-resolved observations of thermal emission are known as thermal phase curves, and show the disk-integrated infrared flux emitted from a planet as it orbits its host star \citep{2008ApJ...678L.129C,2016Natur.532..207D}, revealing the longitudinal structure of its brightness temperature. Eclipse maps show the two-dimensional structure of brightness temperature on a planet's day side \citep{2012ApJ...747L..20M}. 

Atmospheric temperatures and atmospheric circulation are closely related. This is not only due to the ability of an atmosphere to transport heat \citep{2016ApJ...825...99K}, and thus maintain a temperature structure against radiative heating and cooling \citep{2013cctp.book..277S}, but also through the pressure gradient forces in the horizontal momentum equations. These pressure gradient forces mean that any distribution of winds must be balanced by a particular temperature distribution. The most familiar example of such a balance is the geostrophic thermal wind relation \citep[][Chapter 3]{2004dymet.book....H} \begin{equation}
    f\frac{\partial u}{\partial\ln p}=\frac{R}{a}\frac{\partial T}{\partial\vartheta};\quad f\frac{\partial v}{\partial\ln p} = -\frac{R}{a\cos\vartheta}\frac{\partial T}{\partial \lambda};
\end{equation}
which can be derived by assuming the flow is in hydrostatic and geostrophic balance. Above, $p$ is pressure, $a$ is the planetary radius, $R$ is the specific gas constant, and $f=2\Omega\sin\vartheta$ is the Coriolis parameter, with $\Omega$ the planetary rotation rate. $\lambda$ is longitude and $\vartheta$ is latitude. The thermal wind relation states that vertical variation in the horizontal wind $\boldsymbol{u}=(u,v)$ must be accompanied by horizontal variation in the temperature $T$. The utility of geostrophic balance itself is limited to scenarios where the Coriolis term is dominant in the momentum equations, but the requirement that accelerations in the horizontal momentum equations are balanced by horizontal pressure contrasts is a generic one. We will show how an understanding of how different circulation features are balanced by atmospheric temperature structure aids interpretation of observations of thermal emission from extrasolar planets. 

Tidally locked planets are heated on their day side and cool from their night side, and this pattern of heating and cooling strongly influences their atmospheric circulation \citep[see reviews by][]{2013cctp.book..277S,2019AnRFM..51..275P,2020RAA....20...99Z}. Heating on the day side drives divergent motion \citep{2013cctp.book..277S}, which in turn generates stationary Rossby waves (\citealp{2010GeoRL..3718811S}; see also \citealp{1988JAtS...45.1228S}). Interaction between these two circulation components can lead to momentum convergence towards the equator, and the acceleration of a westerly superrotating jet \citep{2011ApJ...738...71S}.


In \citet{2021PNAS..11822705H}, we showed that each of these circulation components can be isolated from one another by separating the horizontal velocity $\boldsymbol{u}$ into a rotational (`divergence free') component $\boldsymbol{u}_{\text{r}}$ and a divergent (`vorticity free') component $\boldsymbol{u}_{\text{d}}$ (called a \emph{Helmholtz decomposition}) \begin{align}
    \boldsymbol{u} &= \boldsymbol{u}_{\text{r}}+\boldsymbol{u}_{\text{d}}, \nonumber \\ 
                   &=  \mathbf{k}\times\nablap\psi + \nablap\chi. \label{eq:rotdiv}
\end{align}
The zonal-mean zonal jet and stationary Rossby waves are contained within the zonal-mean and eddy components of $\boldsymbol{u}_{\text{r}}=\overline{\boldsymbol{u}}_{\text{r}}+\boldsymbol{u}_{\text{r}}^{\prime}$ respectively, and the divergent circulation (comprising, e.g., thermally-direct overturning; cf. \citealp{2021PNAS..11822705H}, and Kelvin waves; see Appendix \ref{sec:sw}) is contained within $\boldsymbol{u}_{\text{d}}$. Above, $\chi$ is the velocity potential and $\psi$ is a streamfunction, defined in terms of the divergence $\delta=\nablap\cdot\boldsymbol{u}$ and vorticity $\zeta=\mathbf{k}\times\nablap\boldsymbol{{u}}$ via the relations $\nablap^{2}\chi=\delta$ and $\nablap^{2}\psi=\zeta$. $\nabla_{p}$ is the horizontal gradient operator acting on isobaric surfaces, and $\mathbf{k}$ is the unit vector in the vertical direction. 

In this paper, our objective is to use the Helmholtz decomposition to understand how different components of atmospheric circulation contribute to the distribution of atmospheric temperature, and by extension thermal emission, on tidally locked planets. This is achieved by decomposing the geopotential $\phi=gz$ into a component that is balanced by purely rotational winds (`the rotational geopotential'), and a component that is due to divergent winds and rotational--divergent interactions (`the divergent geopotential') (Section \ref{sec:height}; see also \citealp{1988JAtS...45.2949T}). We then use the hydrostatic relation \begin{equation}
    \frac{\partial\phi}{\partial\ln p} = -RT 
\end{equation}
to split the global temperature field into rotational and divergent components (Section \ref{sec:temp}). Finally, we show how each circulation component contributes separately to simulated thermal phase curves by linearly expanding $F=\sigma T^{4}$ around the horizontally averaged temperature $T_0$ (Section \ref{sec:olr}). A summary of our results is included at the end of the manuscript (Section \ref{sec:summary}).

The rotational and divergent geopotential components are defined by assuming different balances of terms in the divergence equation (Equation \ref{eq:div1}; introduced in Section \ref{sec:height}). Appendix \ref{sec:sw} illustrates the relationship between balances of terms in the divergence equations, and balances of terms in decomposed momentum equations for the rotational and divergent winds, using the simple example of the linearised shallow water equations. By deriving momentum equations for the rotational and divergent velocities, we show that a steady linear Kelvin wave can exist on a sphere in the absence of drag, in contrast to the equatorial beta-plane where drag is a condition for a linear Kelvin wave \citep{2011ApJ...738...71S}.

In each section, the diagnostics are applied to idealised general circulation model (GCM) simulations of atmospheric circulation on dry, temperate terrestrial tidally locked planets. The model is constructed using the \texttt{Isca} modeling framework \citep{2018GMD....11..843V}, and is described in detail in Appendix \ref{sec:appendix}. We use results from four simulations with different rotation periods, $P=4$, $8$, $16$, and $\infty\ \text{days}$ (i.e., zero rotation). The zero-rotator is included to illustrate the effect of rotation in the other simulations. We also analysed output from a simulation with $P=32\ \text{days}$. The results from this analysis were nearly identical to those for the $P=16\ \text{days}$ simulation and so are not shown. 

In this work, we only present results from simulations of terrestrial planets. However, there is no reason why the techniques described herein cannot be applied to gaseous planets (e.g., `hot Jupiters' or `mini Neptunes') and this is something we aim to do in the near future. 

\begin{figure*}
    \centering\includegraphics[width=.975\textwidth]{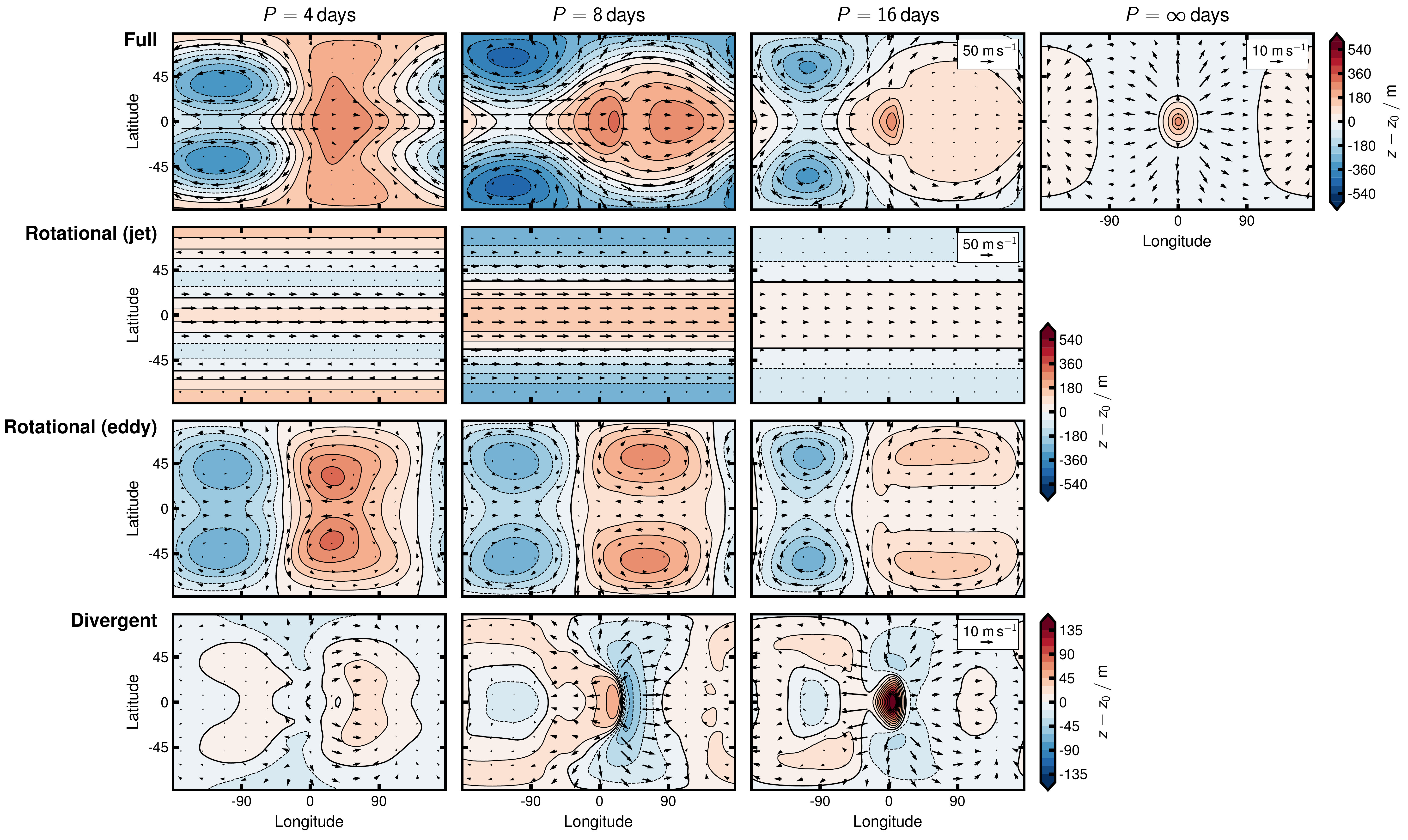}
    \caption{Geopotential height (contours) and horizontal velocity (quivers) at $p=480\,\text{hPa}$ for each GCM simulation. For each simulation, full fields are shown in the top row. The next three rows show the zonal-mean rotational circulation, eddy rotational circulation, and  divergent circulation. Note that the circulation in the $P=\infty\ \text{days}$ simulation has no rotational component.} \label{fig:uz480}
\end{figure*}

\section{Rotational and divergent geopotential components}\label{sec:height}

\subsection{Definition of rotational geopotential}

The horizontal momentum equations can be written in the form \citep[][Chapter 4]{2004dymet.book....H} \begin{equation}
    \frac{\partial\boldsymbol{u}}{\partial t}+\omega\frac{\partial\boldsymbol{u}}{\partial p} + \left(f+\zeta\right)\mathbf{k}\times\boldsymbol{u}=-\nablap\left(\phi+\frac{1}{2}\left\lvert\boldsymbol{u}\right\rvert^{2}\right) -k_{\text{f}}\boldsymbol{u}, \label{eq:mom}
\end{equation}
where $\lvert\boldsymbol{u}\rvert^{2}=u^{2}+v^{2}$, $\omega=\text{D}p/\text{D}t$ is the pressure vertical velocity, and $f=2\Omega\sin\vartheta$ is the Coriolis parameter. $k_{\text{f}}$ is a linear drag coefficient that is non-zero close to the surface. 
Equations for the vorticity $\zeta$ and divergence $\delta$ can then be obtained by taking $\mathbf{k}\cdot\nablap\times\eqref{eq:mom}$ and $\nablap\cdot\eqref{eq:mom}$, respectively, \begin{equation}
    \frac{\text{D}\left(f+\zeta\right)}{\text{D}t}+\left(f+\zeta\right)\delta+\mathbf{k}\cdot\left(\frac{\partial\boldsymbol{u}}{\partial p}\times\nablap\omega\right) = -k_{\text{f}}\zeta, \label{eq:vor} 
\end{equation}\vspace*{-.12in}\begin{align}
    \frac{\partial\delta}{\partial t} + \nablap\cdot\left(\omega\frac{\partial\boldsymbol{u}}{\partial p}\right)+\nablap\cdot&\left[\left(f+\zeta\right)\mathbf{k}\times\boldsymbol{u}\right] \label{eq:div1} \\ &=-\nablap^{2}\left(\phi+\frac{\left\lvert\boldsymbol{u}\right\rvert^{2}}{2}\right) -k_{\text{f}}\delta. \nonumber 
\end{align}
The geopotential $\phi$ does not appear in the vorticity equation (Equation \ref{eq:vor}), and thus all of the dynamics that determines $\phi$ is described by the divergence equation (Equation \ref{eq:div1}). 

In order to separate the geopotential into components associated with the rotational and divergent winds, we substitute $\boldsymbol{u}=\boldsymbol{u}_{\text{r}}+\boldsymbol{u}_{\text{d}}$ into Equation \eqref{eq:div1}, which yields 
\begin{widetext}
\begin{equation}
    \frac{\partial\delta}{\partial t}+\nablap\cdot\left[\omega\frac{\partial\left(\boldsymbol{u}_{\text{r}}+\boldsymbol{u}_{\text{d}}\right)}{\partial p}\right]+u_{\text{d}}\beta+J\left(\chi,\nablap^{2}\psi\right)-\nablap\cdot\left[\left(f+\nablap^{2}\psi\right)\nablap\psi\right]=-\nablap^{2}\left[\phi+\frac{\left\lvert\nablap\psi\right\rvert^{2}}{2}+\frac{\left\lvert\nablap\chi\right\rvert^{2}}{2}+J\left(\psi,\chi\right)\right] -k_{\text{f}}\delta \label{eq:div_expand}
\end{equation}
\end{widetext}
where $J(\psi,\chi)=\mathbf{k}\cdot(\nablap\psi\times\nablap\chi)$ is the Jacobian, and $\beta=a^{-1}\partial f/\partial\vartheta$, with $a$ the planetary radius. A derivation of Equation \eqref{eq:div_expand} is given in Appendix \ref{sec:derive}. 

To decompose the geopotential into components associated with the rotational and divergent circulation we expand it as $\phi=\phi_0+\phi_{\text{r}}+\phi_{\text{d}}$, where $\phi_0(t,p)$ is the horizontal-mean geopotential, which does not contribute to Equation \eqref{eq:div_expand} as $\nablap^{2}\phi_0=0$. We then \emph{define} the `rotational geopotential' as that which is balanced by purely rotational circulation (with no vertical motion), whence we obtain \begin{equation}
    \nablap^{2}\phi_{\text{r}} \equiv \nablap\cdot\left[\left(f+\nablap^{2}\psi\right)\nablap\psi\right] - \nablap^{2}\left(\frac{\left\lvert\nablap\psi\right\rvert^{2}}{2}\right). \label{eq:rotz}
\end{equation}
This equation is called the \emph{non-linear balance equation} \citep{1960Tell...12..364L} and takes the place of the divergence equation for purely non-divergent horizontal circulation (see, e.g., \citealp{2004dymet.book....H}, Chapter 11). For stationary circularly-symmetric flow, it is equivalent to the gradient wind approximation. If $\psi$ (i.e., the rotational wind distribution) is known, then Equation \eqref{eq:rotz} can be inverted to obtain $\phi_{\text{r}}$ (e.g., by expressing Equation \ref{eq:rotz} in terms of spherical harmonics). 

\subsection{Divergent geopotential}

Given the rotational geopotential as defined by Equation \eqref{eq:rotz}, we define the `divergent geopotential' to be that obtained as a residual from \begin{equation}
    \phi_{\text{d}}\equiv\phi - (\phi_0+\phi_{\text{r}}).\label{eq:divz}
\end{equation} 
The remaining terms in Equation \eqref{eq:div_expand} which determine $\phi_{\text{d}}$ involve terms that are purely divergent, and terms due to interaction between the rotational and divergent winds.

\begin{figure*}
    \centering\includegraphics[width=.925\textwidth]{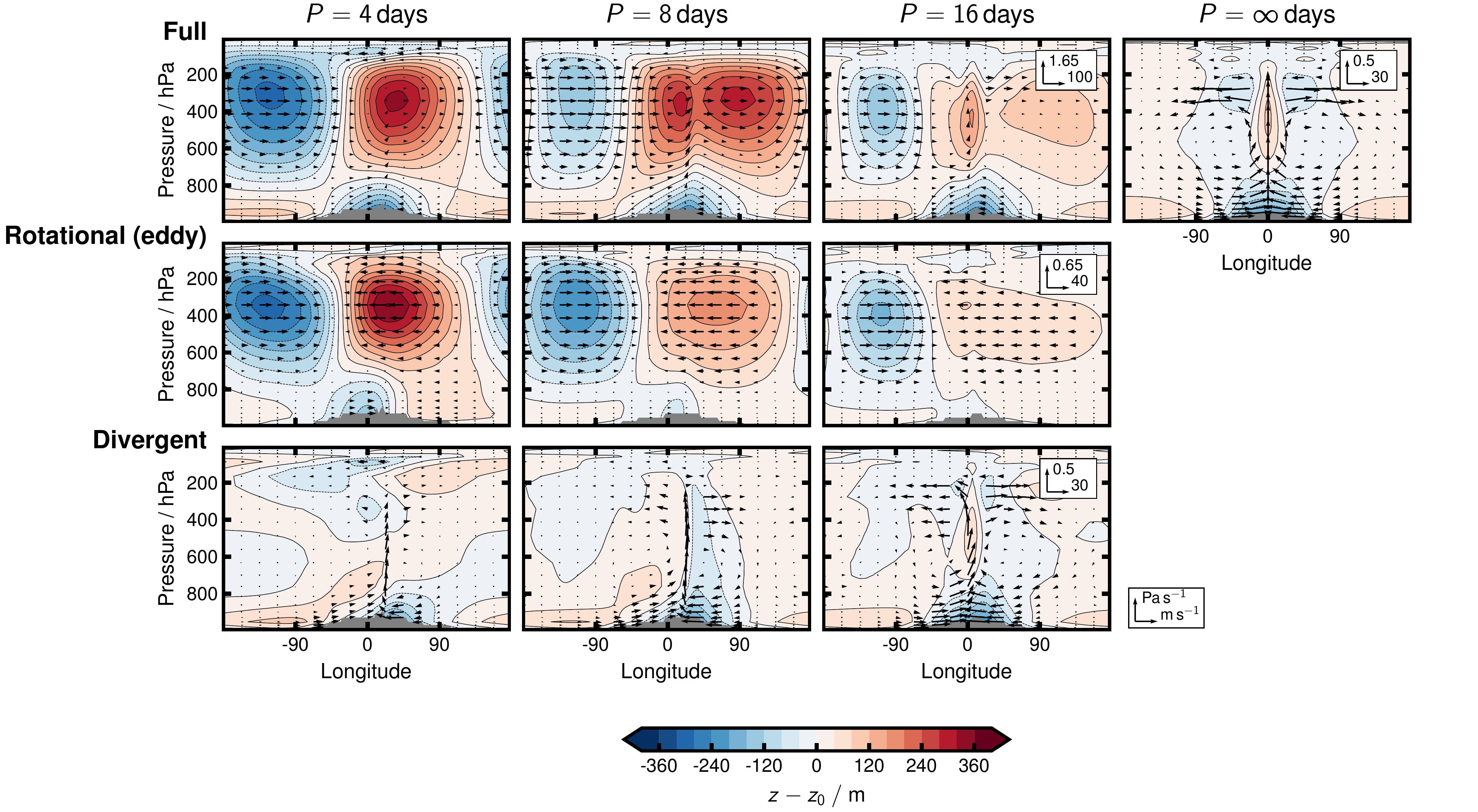}
    \caption{Geopotential height (contours) and velocity ($u$, $\omega$; quivers) in the longitude--pressure plane, averaged between $\pm40^{\circ}$ latitude, shown for each GCM simulation. The top row shows the full circulation. Contributions to the full circulation from the eddy-rotational and divergent components are shown in the next two rows. Data is averaged over the tropics only in order to show the baroclinic structure of the eddy rotational wind.} \label{fig:uzlonp}
\end{figure*}

\subsection{Note on the relationship between balances in the divergence equation and the momentum equations}

In the next subsection we will use Equations \eqref{eq:div_expand}, \eqref{eq:rotz}, and \eqref{eq:divz} to partition the geopotential in our GCM simulations into rotational and divergent components, and will discuss the terms in Equations \eqref{eq:div_expand} and \eqref{eq:rotz} that contribute the most to the decomposed height fields. In anticipation of this, we note that a large term in, e.g., Equation \eqref{eq:rotz}, which defines $\phi_{\text{r}}$, does not \emph{necessarily} correspond to a dominant balance in the momentum equation for $\boldsymbol{u}_{\text{r}}$. This is because an additional term is introduced if Equation \eqref{eq:rotz} is integrated to obtain an equation for $\boldsymbol{u}_{\text{r}}$; the same applies if Equation \eqref{eq:div_expand} (minus the solely rotational terms) is integrated to obtain an equation for $\boldsymbol{u}_{\text{d}}$. This is illustrated for the linearised shallow water equations on the sphere in Appendix \ref{sec:sw}.

\subsection{Application to simulations}

Figure \ref{fig:uz480} shows the horizontal wind $\boldsymbol{u}=(u,v)$ and geopotential height $z$ in the mid-troposphere ($p=480\,\text{hPa}$) for each GCM simulation. The full circulation is shown in the top row, and subsequent rows show the rotational circulation -- split into contributions from the zonal-mean (jet) and eddies (stationary waves) -- and the divergent circulation. Figure \ref{fig:uzlonp} shows the vertical structure of the full circulation, eddy rotational circulation, and divergent circulation averaged between $\pm40^{\circ}$ latitude. In the $P=4\ \text{days}$ simulation, the circulation is mostly contained within the rotational component, whereas in the $P=8$ and $P=16\ \text{days}$ simulations it is more evenly divided into rotational and divergent contributions. Finally, the circulation in the $P=\infty\ \text{days}$ simulation is purely divergent. In this subsection, we describe the important features of each component of the circulation that can be identified in Figures \ref{fig:uz480} and \ref{fig:uzlonp}. We refer to Appendix \ref{sec:decompose} throughout, which contains additional analysis of the contributions of individual terms in the expanded divergence equation (Equation \ref{eq:div_expand}) to each of the rotational and divergent height fields.

The zonal-mean rotational height field $\overline{\phi}_{\text{r}}$ in each of the $P=4$, $8$, and $16\ \text{days}$ simulations is in approximate geostrophic balance with the zonal-mean rotational wind $\overline{u}_{\text{r}}$: \begin{equation}
    f\overline{u}_{\text{r}}\approx-\frac{1}{a}\frac{\partial\overline{\phi}_{\text{r}}}{\partial\vartheta},
\end{equation}
which gives rise to a local height maximum centred on the equator, associated with the superrotating equatorial jet \citep{2018ApJ...869...65H}. In the $P=4\ \text{days}$ simulation there are additional retrograde jets poleward of $\pm50^{\circ}$ in each hemisphere which lead to a secondary maximum in the zonal-mean rotational height field at each pole. 

\begin{figure*}
    \centering\includegraphics[width=.975\textwidth]{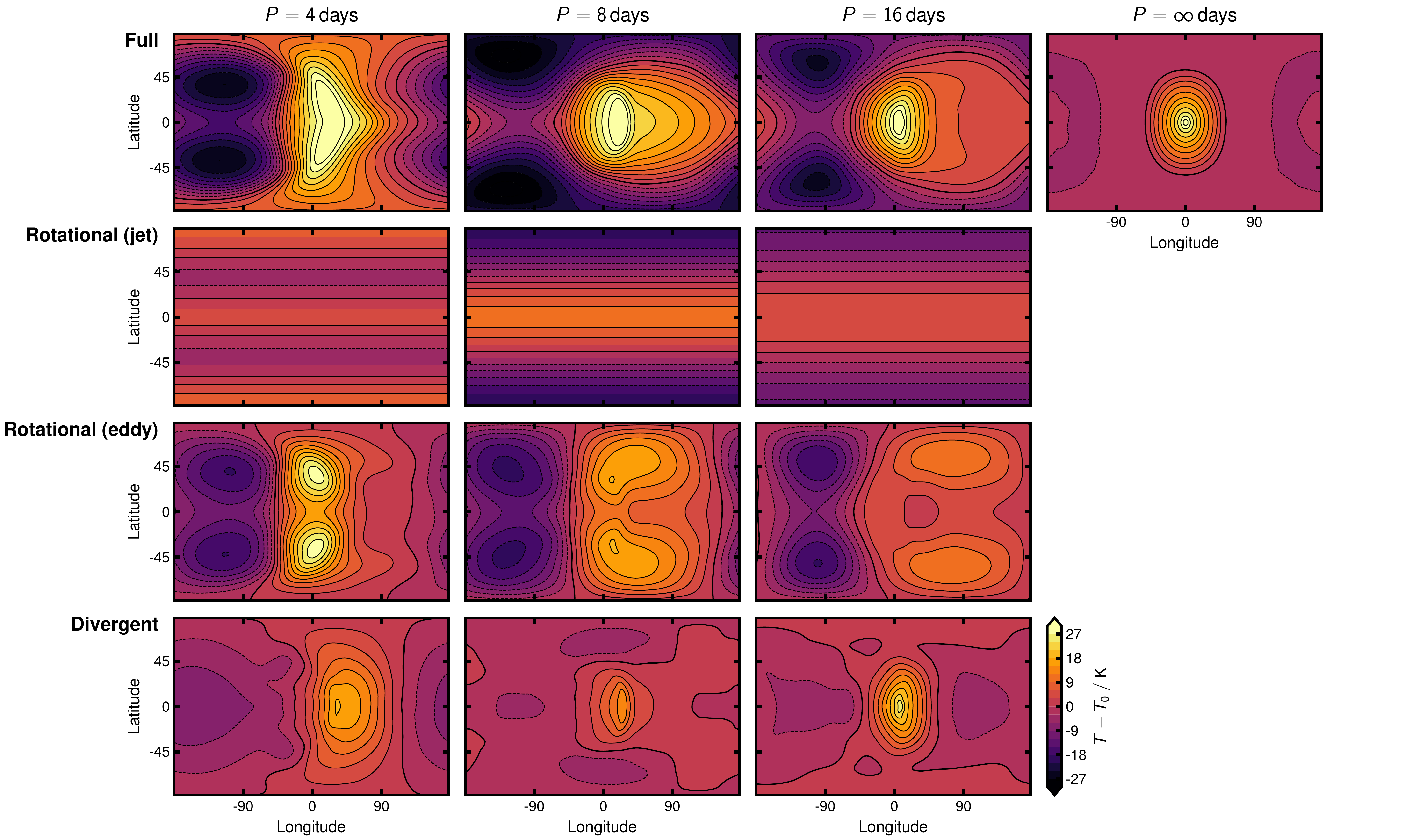}
    \caption{Temperature at $p=750\,\text{hPa}$ for each GCM simulation. For each simulation, full fields are shown in the top row. The next three rows show the zonal-mean rotational circulation, eddy rotational circulation, and  divergent circulation, respectively. } \label{fig:temp750}
\end{figure*}

The eddy-rotational circulation in each simulation has a horizontal wind and geopotential structure characteristic of a stationary equatorial Rossby wave. The longitude of the eddy rotational height maximum is located eastwards of the substellar point due to a Doppler-shift by the zonal-mean zonal wind \citep{2014ApJ...793..141T}. For the rapidly rotating $P=4\ \text{days}$ case, the structure and phase offset of the Rossby waves is very similar to that obtained in a linearised shallow water model with a prescribed background jet (e.g., Figure 10 in \citealp{2018ApJ...869...65H}). As the rotation period is reduced, the Rossby lobes with a positive geopotential anomaly become elongated with respect to those with a negative geopotential anomaly. Similar phenomenology is also exhibited in the GCM simulations presented in \citet[][see their Figure 5]{2011Icar..212....1E}. In Appendix \ref{sec:decompose} we show that this is due to non-linear effects by partitioning the rotational height into contributions from the linear and non-linear terms in the non-linear balance equation (Equation \ref{eq:rotz}). In the vertical, the stationary Rossby waves have a wavenumber-1 modulation \citep{2014ApJ...793..141T}, as can be identified in the second row of Figure \ref{fig:uzlonp} (most clearly evident in the $P=4$ and $8\ \text{days}$ simulations). 

The divergent circulation generally displays the most complex structure of all of the circulation components in Figures \ref{fig:uz480} and \ref{fig:uzlonp}. The exception to this is the $P=\infty\ \text{days}$ simulation, for which the circulation is purely divergent and takes the form of a single isotropic, thermally-direct, overturning cell that features air rising on the day side, near the substellar point, and sinking on the night side. The geopotential structure associated with the overturning circulation is predominantly associated with the $\nablap\cdot(\omega\partial \boldsymbol{u}_{\text{d}}/\partial p)$ term in Equation \eqref{eq:div_expand}, which is dominant in regions of ascent and descent. In the boundary layer the surface friction term is important, and in the outflow regions aloft the horizontal non-linear $-\nablap^{2}[\lvert\nablap\chi\rvert^{2}/2]$ term is important (see Appendix \ref{sec:decompose}).

\begin{figure*}
    \centering\includegraphics[width=.925\textwidth]{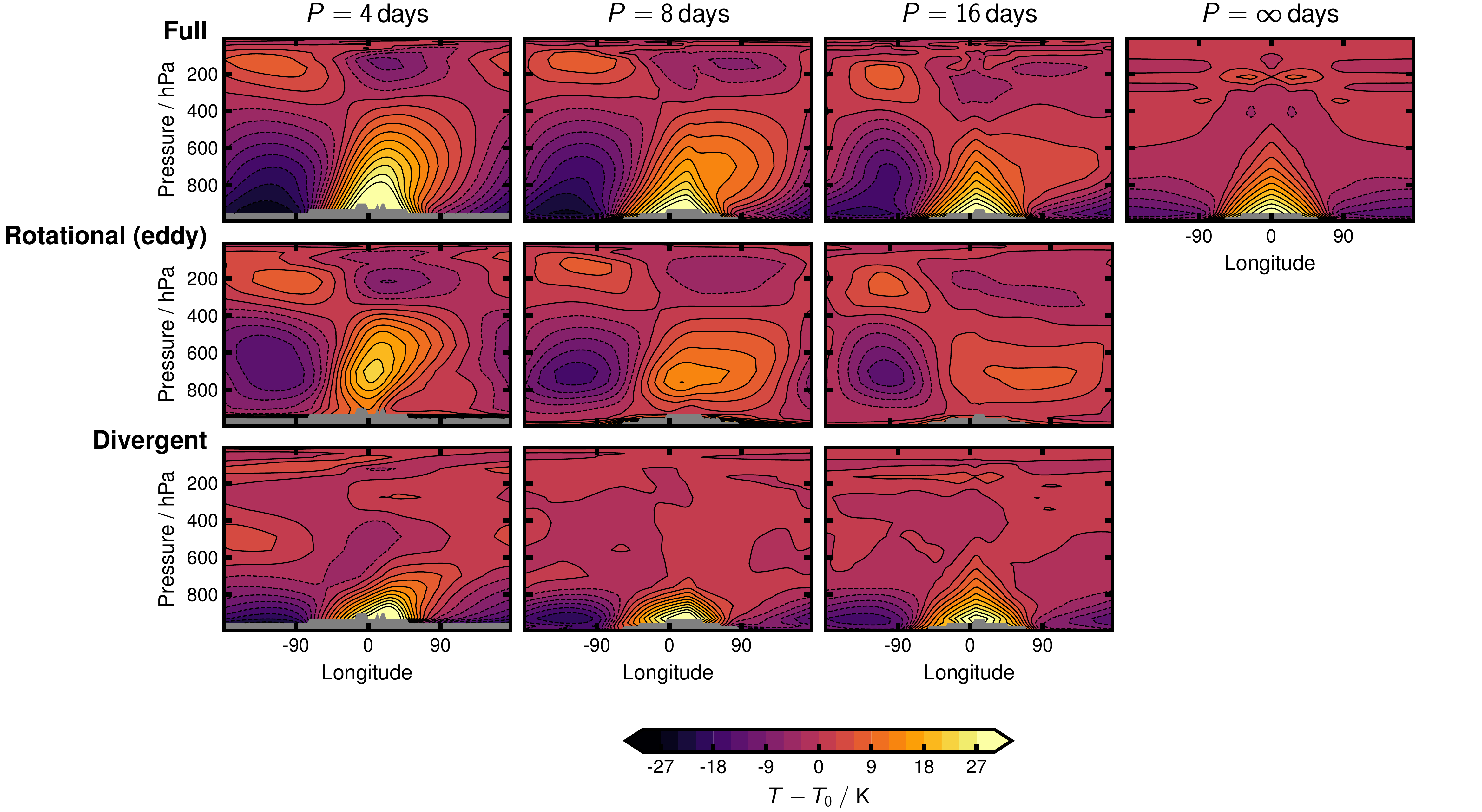}
    \caption{Temperature (contours) in the longitude--pressure plane, averaged meridionally from pole to pole ($\pm90^{\circ}$ latitude), shown for each GCM simulation. The top row shows the full circulation. Contributions to the full circulation from the eddy-rotational and divergent components are shown in the next two rows. }\label{fig:templonp}
\end{figure*}

The balance of terms in Equation \eqref{eq:div_expand} for the $P=16\ \text{days}$ is similar to that in the zero-rotation simulation. In particular, the $\nablap\cdot(\omega\partial \boldsymbol{u}_{\text{d}}/\partial p)$ term is still important in these simulations in the region of strong ascent near the substellar point. However, additional terms in Equation \eqref{eq:div_expand} that describe interactions between the rotational (mostly  $\overline{u}_{\text{r}}$) and divergent circulations are also important (Appendix \ref{sec:decompose}), and our interpretation is that in the $P=16\ \text{days}$ case the overturning circulation is `tipped over' by the eastward superrotating jet.

In the $P=4\ \text{days}$ simulation, the structure of the divergent height field is very different from that in the $P=16$ or $P=\infty\ \text{days}$ simulations, and is harder to interpret. Both its horizontal and vertical structure are somewhat wave-like, although very different to the structure of a linear Kelvin wave on the equatorial beta plane \citep[see, e.g.,][]{1966JMeSJ..44...25M}. In Appendix \ref{sec:decompose} we show that the total divergent height structure shown in Figures \ref{fig:uz480} and \ref{fig:uzlonp} for the $P=4\ \text{days}$ simulation is mostly due to the linear $u_{\text{d}}\beta$ term, which takes the form of a Kelvin wave in a linearised shallow water model (Appendix \ref{sec:sw}), and terms describing interaction between rotational and divergent horizontal winds. Finally, in the $P=8\ \text{days}$ simulation, the divergent height field has features that are characteristic of both the slowly rotating simulations, with thermally-direct overturning, and the $P=4\ \text{days}$ simulation with non-linear and wave-like contributions to the height, suggesting that it resides in an intermediate regime.

\newpage

\section{Relationship between geopotential components and temperature}\label{sec:temp}

In this section, we use the hydrostatic relation to obtain components of the atmospheric temperature that are associated with the rotational and divergent geopotential components. This is achieved by requiring that each component of the geopotential ($\phi_0$, $\phi_{\text{r}}$ and $\phi_{\text{d}}$) is in hydrostatic balance with the corresponding component of the temperature ($T_0$, $T_{\text{r}}$ and $T_{\text{d}}$), i.e., \begin{equation}
    \frac{\partial\phi_0}{\partial\ln p} \equiv -RT_0;\quad 
    \frac{\partial\phi_{\text{r}}}{\partial\ln p} \equiv -RT_{\text{r}};\quad 
    \frac{\partial\phi_{\text{d}}}{\partial\ln p} \equiv -RT_{\text{d}}.
\end{equation}
Understanding how each component of the circulation contributes to atmospheric temperature structure is important for interpreting observations of thermal emission from tidally locked planets, such as eclipse maps \citep{2012ApJ...747L..20M} and phase curves \citep{2008ApJ...678L.129C,2016Natur.532..207D,2021arXiv211011837M}.

Figure \ref{fig:temp750} shows the temperature structure in the lower troposphere ($p=750\,\text{hPa}$) for each GCM simulation. As with the geopotential height shown in Figure \ref{fig:uz480}, the temperature is decomposed into contributions from the zonal-mean and eddy rotational circulations, and the divergent circulation. Figure \ref{fig:templonp} shows the vertical structure of the full circulation, eddy rotational circulation, and divergent circulation, averaged meridionally over all latitudes for consistency with the disk-integrated phase curves shown later.

\begin{figure*}
    \centering\includegraphics[width=\textwidth]{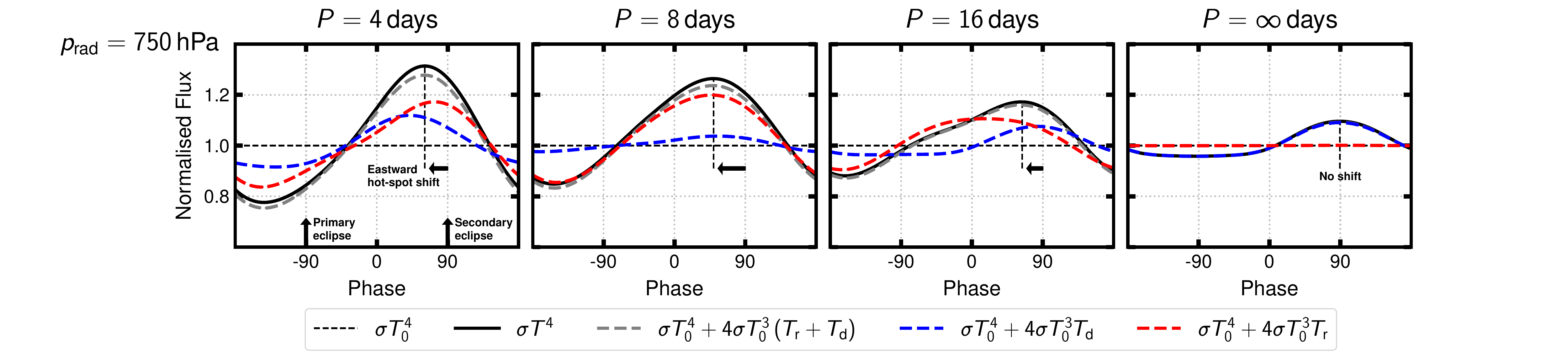}
    \caption{Phase curves calculated using the broadband thermal emission from a radiating level at $p_{\text{rad}}=750\,\text{hPa}$, for each GCM simulation. The secondary and primary eclipses would be at phase $90^{\circ}$ and $-90^{\circ}$. The solid black curve shows emission calculated using the full temperature field, and the dashed grey curve shows emission using the linear approximation given by Equation \eqref{eq:lin_emission}. The dashed red and blue curves show the contributions to the linear approximation from the rotational and divergent temperature fields, respectively.}\label{fig:phase}
\end{figure*}

Generally, the horizontal and vertical structures of the temperature components shown in Figures \ref{fig:temp750} and \ref{fig:templonp} are similar to the structure of the height components shown in Figures \ref{fig:uz480} and \ref{fig:uzlonp}. The main different between them is a phase offset in their vertical structures, which appears as the temperature is obtained from the \emph{vertical derivative} of the geopotential (this is why the temperature in Figure \ref{fig:temp750} is plotted deeper in the atmosphere than the height field in Figure \ref{fig:uz480}).

In each of the simulations with rotation, there is an eastward `hot spot' shift from the substellar point. In the $P=8$ and $16\ \text{days}$ simulations, this is almost exclusively due to the rotational standing Rossby waves which are Doppler-shifted eastwards from the substellar point by the zonal-mean jet \citep{2014ApJ...793..141T,2018ApJ...869...65H}. The zonal-mean rotational temperature that balances the zonal-mean rotational wind cannot contribute a longitudinal shift to the temperature structure by definition, although it does contribute towards `brightening' the equator with respect to the poles. 
The divergent temperature structure in the $P=16\ \text{days}$ simulation is similar to that of the $P=\infty\ \text{days}$ simulation which only features overturning circulation. The overturning circulation is a direct response to heating at the substellar point, and it appears that this strong coupling to the surface prevents it from being Doppler-shifted much by the jet.

By contrast, in the $P=4\ \text{days}$ simulation, the divergent temperature structure has a \emph{greater} eastward shift than the eddy rotational circulation (which is now only shifted by a few degrees longitude from the substellar point), although its amplitude is comparatively lower. The eastward shift of the divergent temperature in this simulation is consistent with the wave-like nature of its divergent circulation and height field, discussed in the previous section. The temperature structure associated with the zonal mean rotational circulation is also different to that in the $P=8$ and $16\ \text{days}$ simulations, due to the retrograde jets that exist in this simulation in mid-latitudes, and now contributes a brightening to both the equator and the poles, relative to the mid-latitudes.

\section{Thermal phase curves}\label{sec:olr}

Thermal phase curves show the infrared flux emitted from a planet as it orbits its star. As a planet progresses through its orbit we see emission from different angles, so these observations characterise the longitudinal distribution of temperature on extrasolar planets \citep{2008ApJ...678L.129C, 2015ApJ...802...21K}. Thermal phase curves are already available for hot Jupiters and high-temperature rocky planets \citep{2016Natur.532..207D,2019Natur.573...87K}, and should increase in quality in the coming years \citep{2018PASP..130k4402B} following the successful launch of the \emph{James Webb Space Telescope}. With \emph{JWST} it should also be possible to produce thermal phase curves for smaller and cooler terrestrial planets \citep{2009PASP..121..952D}. 

In this section, we compute thermal phase curves from the decomposed temperature structures to show how each component contributes to their observable features. We approximate that the broadband thermal emission $I^{\uparrow}$ originates from a specific radiating level $p_{\text{rad}}$, so that \begin{equation}
    I^{\uparrow}\left(p\,{=}\,p_{\text{rad}}\right) = \sigma T^{4}\left(p\,{=}\,p_{\text{rad}}\right). \label{eq:nonlin_emission}
\end{equation}
We do this to simulate observing at a wavelength at which the atmosphere is optically thick, instead of using the model's actual semi-grey outgoing longwave radiation (OLR), because in our simulations the OLR is dominated by the surface emission and so is not an instructive demonstration of the potential contribution to the phase curve from the atmosphere. This also corresponds to the case of gaseous planets where the thermal emission is entirely due to the atmosphere and its circulation.

As Equation \eqref{eq:nonlin_emission} is non-linear in $T$, we cannot easily separate out contributions to $I^{\uparrow}$ from each of $T_{\text{r}}$ and $T_{\text{d}}$. We therefore linearise Equation \eqref{eq:nonlin_emission} about the horizontal mean temperature $T_0$, \begin{equation}
    I^{\uparrow} \approx \sigma T^{4}_{0} + 4\sigma T^{3}_{0}\left(T - T_{0}\right) = \sigma T^{4}_{0} +4\sigma T^{3}_{0}\left(T_{\text{r}} + T_{\text{d}}\right),\label{eq:lin_emission}
\end{equation}
which separates contributions to $I^{\uparrow}$ from $T_{\text{r}}$ and $T_{\text{d}}$. Normalised thermal phase curves are then computed according to \begin{equation}
    I\left(\xi\right) = \frac{1}{\pi\sigma T_0^{4}}\int^{-\xi+\nicefrac{\pi}{2}}_{-\xi-\nicefrac{\pi}{2}}\int^{\nicefrac{\pi}{2}}_{\nicefrac{-\pi}{2}}I^{\uparrow}\cos\left(\lambda+\xi\right)\cos^{2}\vartheta\,\text{d}\vartheta\,\text{d}\lambda.\label{eq:phase}
\end{equation}
where $\xi$ is the phase angle of the planet in its orbit \citep{2008ApJ...678L.129C}.

Figure \ref{fig:phase} shows phase curves computed for each GCM simulation using Equation \eqref{eq:phase}, for a radiating level located at $p=750\,\text{hPa}$, which corresponds to the level for which the horizontal temperature structure was shown in Figure \ref{fig:temp750}. Solid black curves show the phase curve computed using the non-linear expression given by Equation \eqref{eq:nonlin_emission}, and dashed grey curves show the phase curve computed using the linear approximation given by Equation \eqref{eq:lin_emission}. There is generally good agreement between the phase curves computed using the non-linear and linear expressions, and the difference between the two is reduced with increasing rotation period as deviations from the horizontal-mean temperature become smaller. This shows that the decomposition of the phase curve into rotational and divergent components by the linear approximation is useful.

The contributions from the rotational and divergent components of the temperature to the linear approximation are shown in Figure \ref{fig:phase} as dashed red and blue lines, respectively. In each simulation with rotation, the amplitude of the rotational contribution to the phase curve is larger than that of the divergent contribution, and leads to a negative offset in the phase curve peak from $\xi=90^{\circ}$ (corresponding to an eastward hot-spot shift in the temperature maps shown in Figure \ref{fig:temp750}). As discussed in Section \ref{sec:temp}, this is entirely due to the Doppler-shifted Rossby waves, as the temperature structure that balances the jet cannot itself contribute a direct hot-spot shift by definition (although it is the velocity of the jet that Doppler-shifts the waves). 

\begin{figure}
    \centering\includegraphics[width=.4\textwidth]{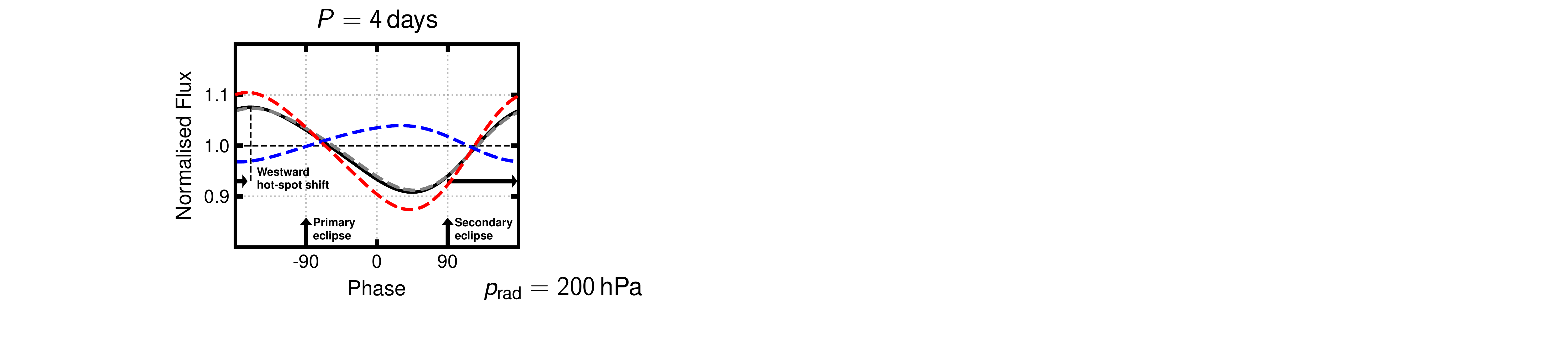}
    \caption{Phase curve calculated using broadband thermal emission from a radiating level at $p=200\,\text{hPa}$ for the $P=4\ \text{days}$ simulation. See the caption of Figure \ref{fig:phase} for the meaning of each curve.}\label{fig:phase2}
\end{figure}

The negative phase offset associated with the stationary Rossby waves increases monotonically with increasing period. However, this trend is not reflected in the full phase curves, due to contributions from the divergent circulation. In the $P=4\ \text{days}$ simulation, the divergent circulation, which in this simulation resembles a non-linear wave that is Doppler-shifted eastwards, also contributes an eastward phase shift, that is in fact larger than that associated with the rotational circulation. This means that the peak of the full phase curve has a greater negative phase offset from $\xi=90^{\circ}$ than it would if only the rotational circulation (stationary Rossby waves) contributed. By contrast, in the $P=16\ \text{days}$ simulation, the divergent circulation now takes the form of overturning circulation, and is restricted to be close to the substellar point, due to strong coupling between the overturning circulation and the instellation. This means that the divergent contribution to the phase curve is not offset from $\xi=90^{\circ}$, and the full phase curve has a lesser negative phase offset than it would if due to the stationary Rossby waves alone. 

Our analysis of the vertical temperature structure in the previous section shows that the phase curves in Figure \ref{fig:phase} will depend strongly on our choice of radiating level. To illustrate this we show an additional phase curve in Figure \ref{fig:phase2} for the $P=4\ \text{days}$ simulation, this time for $p_{\text{rad}}=200\,\text{hPa}$. In this figure, the sign of the normalised flux for each curve (full, rotational, and divergent) is essentially flipped with respect to that shown in Figure \ref{fig:phase}, and the peak of the full phase curve has a positive offset from $\xi=0^{\circ}$, which would correspond to a westward hot-spot shift in temperature. In our simulation, this is due to the wavenumber-1 structure of the stationary Rossby waves in the vertical (see Figure \ref{fig:templonp}), which means that they can contribute either a negative or a positive shift to the phase curve peak depending on the exact pressure level from which they contribute to thermal emission. This is a key result of understanding thermal phase shifts as due to stationary wave shifts rather than heat advection by a jet, and is a potential explanation for unexpected observations of westward phase shifts \citep{2018NatAs...2..220D}.

\vfill

\section{Summary} \label{sec:summary}

\citet{2021PNAS..11822705H} showed that the Helmholtz decomposition $\boldsymbol{u}=\boldsymbol{u}_{\text{r}}+\boldsymbol{u}_{\text{d}}$ is a useful tool for separating out different components of the atmospheric circulation on tidally locked planets. The jet and stationary Rossby waves are contained within $\boldsymbol{u}_{\text{r}}$ and divergent circulation 
is contained within $\boldsymbol{u}_{\text{d}}$, which takes the form of thermally-direct overturning for slowly rotating planets (e.g., our $P=8$ and $P=16\ \text{days}$ simulations; see also \citealp{2021PNAS..11822705H}), but has a complex linear Kelvin / non-linear wave structure for more rapidly rotating planets (cf. our $P=4\ \text{days}$ GCM simulation, and the shallow water results in Appendix \ref{sec:sw}). 

The objective of this work was to build upon \citet{2021PNAS..11822705H} by relating the atmospheric temperature structure to each of the circulation components contained within $\boldsymbol{u}_{\text{r}}$ and $\boldsymbol{u}_{\text{d}}$. This was achieved by defining rotational and divergent components of the geopotential using balance relations in the divergence equation, which were then used to define a temperature structure via the hydrostatic relation. To illustrate the utility of decomposing the temperature in this way, we applied it to output from idealised GCM simulations of atmospheric circulation on terrestrial tidally locked planets (run using \texttt{Isca}; \citealp{2018GMD....11..843V}), considering four rotation periods: $P=4$, $8$, $16$, and $\infty\ \text{days}$ (i.e., zero rotation).

The temperature maps we obtain (Figure \ref{fig:temp750}) show that both the rotational and divergent circulations make non-negligible contributions to the horizontal temperature structure. Understanding what determines the relative strength of these circulations, and how they depend on properties of a particular exoplanet, will be crucial for correctly interpreting observations of thermal emission such as phase curves \citep{2008ApJ...678L.129C} and eclipse maps  \citep{2012ApJ...747L..20M}. 

Our analysis shows that both circulation components can contribute to the hot spot shifts that have been inferred from observations of thermal emission (e.g., \citealp{2016Natur.532..207D}) (Figures \ref{fig:temp750} and \ref{fig:phase}). In our simulations, both the rotational circulation and the divergent circulation contribute an eastward hot spot shift if thermal emission is assumed to come from the lower troposphere ($p=750\,\text{hPa}$). For our $P=8$ and $P=16\ \text{days}$ simulations, the hot spot shift is almost entirely due to the rotational stationary Rossby waves (which are Doppler-shifted eastwards by the superrotating jet). In the $P=8\ \text{days}$ case, this is because the amplitude of the rotational temperature component is much larger than the divergent temperature component. In the $P=16\ \text{days}$ case, both components have a similar amplitude, but the divergent component does not contribute a hot spot shift as it is dominated by thermally-driven overturning, which means it is strongly coupled to the pattern of instellation which is centered on the substellar point. By contrast, in the $P=4\ \text{days}$ simulation, both the rotational component and the divergent component (which is now more wave-like) contribute an eastward hot spot shift if emission is from the lower troposphere. However, we show that the rotational component can also contribute a westward hot spot shift if emission comes from higher up in the atmosphere ($p=200\,\text{hPa}$ in our simulations; Figure \ref{fig:phase2}), which is due to its wavenumber-1 structure in the vertical (Figure \ref{fig:templonp}). 

From this analysis, we suggest that the temperature structure of the atmosphere of a tidally locked planet can be divided into four physically meaningful components:

\begin{enumerate}
    \item The overturning divergent component, providing a temperature peak at the substellar point and uniform temperature elsewhere.
    \item A wave-like divergent component, providing a zonal-wavenumber-1 sinusoidal temperature modulation on the equator which can be Doppler-shifted by a zonal jet.
    \item The Rossby-wave-like eddy rotational component, providing a zonal-wavenumber-1 sinusoidal temperature modulation with maxima off the equator, which can be Doppler-shifted by a zonal jet. 
    \item The zonal-mean rotational component, i.e. the zonal jet, providing a zonally uniform temperature field with a meridional gradient determined by the gradient of the jet itself.
\end{enumerate}

These components could be used to fit thermal phase curves or eclipse maps with physically meaningful functions. We hope that the techniques and analysis presented herein will be of use for interpreting observations of thermal emission on tidally locked planets.

\begin{acknowledgements}
    NTL was supported by Science and Technology Facilities Council grant ST/S505638/1. MH was supported by a Junior Research Fellowship at Christ Church, Oxford. 
    The authors are grateful to Dr. Man-Suen Chan for IT support. 
\end{acknowledgements}

\appendix 

\begin{figure*}[!t]
    \centering\includegraphics[width=.8\textwidth]{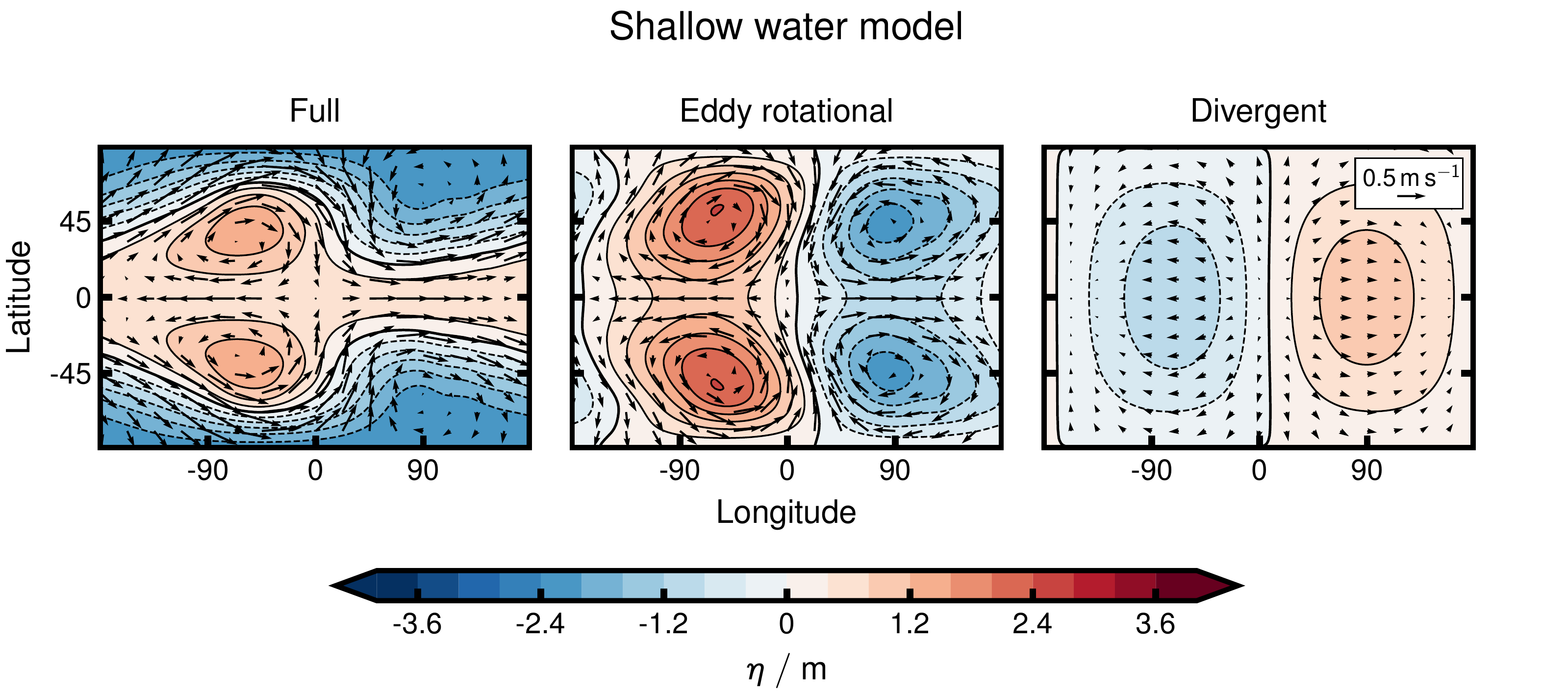}
    \caption{Free surface displacement (height; colour contours) and horizontal velocity (quivers) obtained in a linear shallow water simulation without drag ($P=8\ \text{days}$, Earth's radius). The height and velocity fields are decomposed into eddy rotational (center panel) and divergent (right panel) components. The zonal-mean rotational component is small and so is not shown. }\label{fig:sw}
\end{figure*}

\section{Rotational and divergent circulations in linear shallow water on the sphere}\label{sec:sw}

In this appendix, we use the example of the linearised shallow water equations on the sphere to show how partitioning the height field into rotational and divergent components can be used to obtain momentum equations for the rotational and divergent wind. We use these equations to interpret the existence of a steady linear Kelvin wave in the spherical geometry, in the absence of drag, which does not exist on the equatorial beta plane. 

In the absence of friction, the linearised shallow water equations may be written \citep[][Chapter 3]{2017aofd.book.....V} \begin{align}
    \frac{\partial\boldsymbol{u}}{\partial t} + f\mathbf{k}\times\boldsymbol{u}&=-g\nabla\eta, \label{eq:sw} \\ 
    \frac{\partial\eta}{\partial t} + H\delta &= Q. 
\end{align}
Above, $\eta$ is the free surface displacement (or height), $H$ is the mean fluid depth, and $Q$ is a mass source or sink term.

As with the primitive equations (Section \ref{sec:height}), equations for the vorticity and divergence can be obtained by taking $\mathbf{k}\cdot\nabla\times\eqref{eq:sw}$ and $\nabla\cdot\eqref{eq:sw}$, respectively,  \begin{align}
    \frac{\partial\zeta}{\partial t} + f\delta + \left(v_{\text{r}}+v_{\text{d}}\right)\beta &= 0, \\ 
    \frac{\partial\delta}{\partial t} - \nabla\cdot\left[f\nabla\psi\right]+u_{\text{d}}\beta &= -g\nabla^{2}\eta.
\end{align}
Assuming a steady state, we can partition the divergence equation into two equations that define the rotational and divergent height, $\eta_{\text{r}}$ and $\eta_{\text{d}}$: \begin{equation}
    \nabla\cdot\left[f\nabla\psi\right]\equiv g\nabla^{2}\eta_{\text{r}};\quad -u_{\text{d}}\beta\equiv g\nabla^{2}\eta_{\text{d}},
\end{equation}
which can then be integrated to yield momentum equations for the rotational and divergent circulations in a steady state \begin{align} 
    f\mathbf{k}\times\boldsymbol{u}_{\text{r}}&=-g\left(\nabla \eta_{\text{r}} + \mathbf{k}\times\nabla\gamma\right), \label{eq:swr}\\ 
    f\mathbf{k}\times\boldsymbol{u}_{\text{d}}&=-g\left(\nabla \eta_{\text{d}} - \mathbf{k}\times\nabla\gamma\right), \label{eq:swd}
\end{align} 
where $\gamma$ is defined by \begin{equation}
    g\nabla^{2}\gamma\equiv\beta v_{\text{r}}=-\left(f\delta+\beta v_{\text{d}}\right). \label{eq:gamma}
\end{equation}
Equation \eqref{eq:gamma} can be obtained by taking the curl of either Equation \eqref{eq:swr} or \eqref{eq:swd}, and the second equality comes from the vorticity equation in steady state.

Equations \eqref{eq:swr} and \eqref{eq:swd} each contain an additional term involving the potential function $\gamma$, when compared with the momentum equation for the full horizontal velocity $\boldsymbol{u}$. These terms arise as an integration constant and are due to the variability of $f$ with latitude, and their proper representation is contingent on working in a spherical geometry \citep{1988JAtS...45.2949T}. 

Figure \ref{fig:sw} shows horizontal velocity and height fields from a linear shallow water simulation using the GFDL\footnote{Geophysical Fluid Dynamics Laboratory, Princeton. \url{https://www.gfdl.noaa.gov/idealized-spectral-models-quickstart/}} spectral shallow water model. We ran the model at T85 resolution for 500 days with a timestep of 300 seconds, and plot output averaged over the final 100 days of the run. The mass source term is configured for a tidally locked planet as \begin{equation}
    Q = \frac{\eta_{\text{eq}}-\eta}{\tau_{\text{rad}}}, \label{eq:masssource}
\end{equation}
following \citet{2013ApJ...776..134P}, where $\eta_{\text{eq}}$ is given by \begin{equation}
    \eta_{\text{eq}}=\begin{cases}H+\Delta\eta_{\text{eq}}\cos\lambda\cos\vartheta & \text{on the day side}, \\ 
        H & \text{on the night side}. 
    \end{cases}
\end{equation}
$\Delta\eta_{\text{eq}}$ is the difference in the radiative equilibrium height between the substellar point and the terminator. In Equation \eqref{eq:masssource}, $\tau_{\text{rad}}$ is the radiative relaxation timescale. The GFDL model solves the full non-linear shallow water equations, and the solution was kept in a linear regime by using a small forcing amplitude, $\eta_{\text{eq}}/H\ll1$, following \citet{2013ApJ...776..134P}. An important feature of the model is that it does not include a linear drag (i.e., $\tau_{\text{drag}}=\infty$ in \citealp{2013ApJ...776..134P}). The planetary parameters we used for this simulation were: $a=6.371\times10^{6}\,\text{m}$, $P=8\ \text{days}$, and $g=9.81\,\text{m\,s}^{-2}$. The equilibrium height was $H=10000\,\text{m}$, the magnitude of the forcing was $\Delta\eta_{\text{eq}}=10\,\text{m}$, and the radiative timescale was $\tau_{\text{rad}}=0.1\ \text{days}$.  We chose these parameters for demonstrative purposes as they give rotational and divergent components of similar magnitude, but our conclusions are not sensitive to the chosen parameters as long as the forcing is linear. In Figure \ref{fig:sw} the velocity and height is split into eddy rotational and divergent components. The zonal mean rotational component is not shown as it is small. 

The decomposed rotational and divergent circulations shown in Figure \ref{fig:sw} have structures that resemble the linear solutions for equatorial Rossby and Kelvin waves, respectively, on the equatorial beta plane \citep{1966JMeSJ..44...25M}, although they are modified by the spherical geometry. This partitioning of the Rossby waves into the rotational component and the Kelvin waves into the divergent component is consistent with \citet{1994GApFD..76..121V}, who show that in the limit of small Lamb parameter, $\Lambda\equiv4\Omega^{2}a^{2}/(gH)\ll1$ (in our simulation $\Lambda\approx0.1$), Rossby waves are purely rotational and Kelvin waves are purely divergent. The Kelvin wave obtained here has non-zero meridional velocity, which derives from the spherical geometry \citep{2019Icar..322..103Y}. The Rossby component is very similar to that obtained by \citet{2011ApJ...738...71S} on the equatorial beta plane in the absence of friction (as is the case here; cf. their Figure 14). Notably, \citet[][their Appendix C]{2011ApJ...738...71S} show that in the absence of drag, an equatorial Kelvin wave \emph{cannot exist} on the equatorial beta plane. 

In a steady state, the zonal momentum equation is \begin{equation}
    fv = \frac{g}{a\cos\vartheta}\frac{\partial\eta}{\partial\lambda}. \label{eq:zonal_sw}
\end{equation}
As $f=0$ at the equator (and also $v=0$ due to the hemispheric symmetry of the problem), Equation \eqref{eq:zonal_sw} requires that $\partial\eta/\partial\lambda=0$ at the equator. This is satisfied in our linear shallow water simulation (see the left-hand panel in Figure \ref{fig:sw}). However, \citet{2011ApJ...738...71S} additionally show that on the equatorial beta plane, equatorial Rossby wave solutions have no longitudinal $\eta$ variation at the equator, which by Equation \eqref{eq:zonal_sw}, means that an equatorial Kelvin component cannot exist (as it would introduce a non-zero $\partial\eta/\partial\lambda$). 

\begin{figure}[!t]
    \centering\includegraphics[width=.4\textwidth]{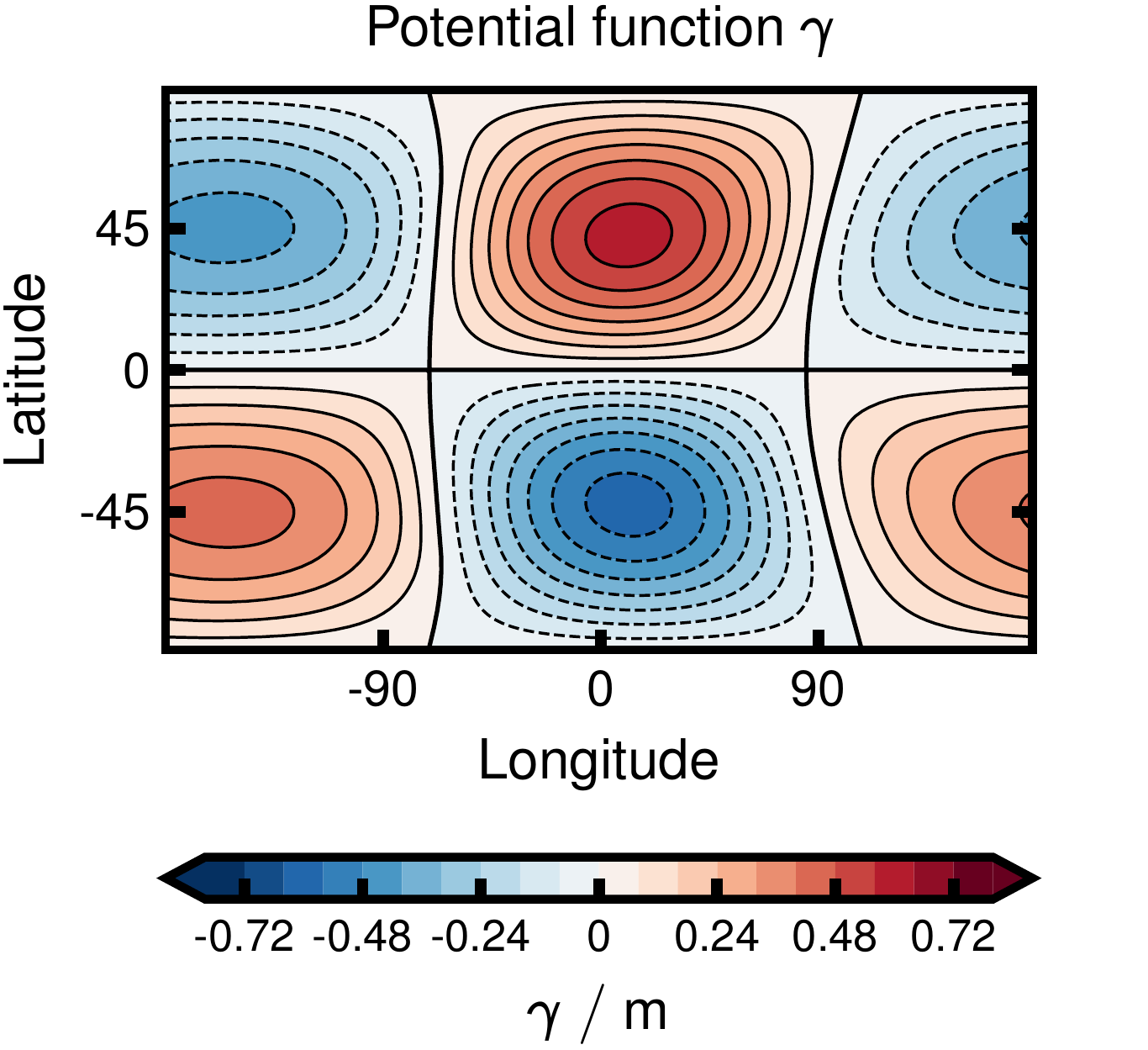}
    \caption{Structure of the potential function $\gamma$ that appears in the momentum equations for the rotational and divergent circulations, obtained from the linear shallow water simulation.}\label{fig:gamma}
\end{figure}

By contrast, both the equatorial Rossby and Kelvin components in Figure \ref{fig:sw} have a non-zero height modulations that cancel one another so that the total $\partial\eta/\partial\lambda$ at the equator is still zero. This is due to the effect of the spherical geometry on the waves, which now are governed by a zonal momentum equation of the form (e.g., for $u_{\text{r}}$, from Equation \ref{eq:swr}) \begin{equation}
    fv_{\text{r}}=g\left(\frac{1}{a\cos\vartheta}\frac{\partial\eta_{\text{r}}}{\partial\lambda}+\frac{1}{a}\frac{\partial\gamma}{\partial\theta}\right)
\end{equation}
If $\partial\gamma/\partial\theta$ is non-zero at the equator (which will be the case if $v$ is symmetric about the equator; $\gamma$ obtained in our simulation is shown in Figure \ref{fig:gamma} for reference), then neither will $\partial{\eta_{\text{r}}}/{\partial\lambda}\lvert_{\vartheta=0}=-\partial{\eta_{\text{d}}}/{\partial\lambda}\lvert_{\vartheta=0}$. As a consequence, on the sphere and in the limit of small Lamb parameter, the existence of a rotational stationary Rossby wave requires the existence of a divergent stationary Kelvin wave, and vice versa. \citet{2011ApJ...738...71S} introduced a linear drag to their beta plane model to generate the stationary Kelvin wave necessary for the acceleration of a zonal jet, but we have shown here that this linear drag is not necessary for the formation of a Kelvin (-like) wave on a sphere rather than on a beta plane.

\section{Description of General Circulation Model}\label{sec:appendix}

The GCM simulations analysed in the main text were run using \texttt{Isca}, which is a a framework for building idealised general circulation models of varying complexity \citep{2018GMD....11..843V}. \texttt{Isca} is built on the GFDL spectral dynamical core, which is used to integrate the primitive equations of motion forwards in time. In the present study we use a dry GCM configuration, i.e., water vapour and moist processes are not included in the model atmosphere. The model is configured to simulate the atmospheric circulation of terrestrial planets with a lower boundary modeled as a `slab ocean'. All planetary parameters aside from the rotation rate and properties of the stellar irradiation (e.g., radius, surface pressure, gravitational acceleration) are set to be those of the Earth. Four rotation periods are used, $P=4$, $8$, $16$, and $\infty\ \text{days}$ (as described in the main text), and the distribution of stellar irradiation is set to be that of a tidally locked planet (defined below).  The main components of the model are described below. 

\subsection{Dynamical core}

The dynamical core integrates the primitive equations forwards in time, using a semi-implicit leapfrog scheme with a Robert--Asselin time filter. The equations are solved in vorticity-divergence form on a thin spherical shell (similar to Equations \ref{eq:vor} and \ref{eq:div1}) using a pseudospectral method in the horizontal (prognostic fields are represented by a triangular truncation of spherical harmonics), and a finite difference method in the vertical (see, e.g., \citealp{satoh2013atmospheric}, Chapters 21--23). The vertical coordinate used is a terrain-following `sigma' coordinate, defined $\upsigma=p/p_{\text{s}}$, where $p$ is pressure and $p_{\text{s}}$ is surface pressure. The simulations presented here use a T42 triangular truncation, which corresponds to approximately $2.8^{\circ}$ latitude--longitude resolution. We use 25 unevenly spaced levels in the vertical, distributed according to $\upsigma=\exp[-5(0.05\tilde{z}+0.95\tilde{z}^{3})]$ where $\tilde{z}$ is equally spaced in the unit interval. A $\nabla^{8}$ hyperviscosity term is applied to the momentum and thermodynamic equations operating on a timescale of $0.1\ \text{days}$ at the grid scale. The data analysed in the main text is interpolated from the model's sigma coordinate to a pure pressure coordinate. 

\subsection{Radiative transfer}

We use a semi-grey radiative transfer code to model radiative heating and cooling. Radiative fluxes are split into two wavelength bands: one covering longwave (thermal emission) wavelengths, and the other covering shortwave (stellar heating) wavelengths. 

In the shortwave band, the atmosphere is assumed to be transparent, and all incoming radiation is absorbed at the surface (i.e., the surface albedo has been folded into the solar constant). The instellation at the top-of-atmosphere is imposed with the following distribution, appropriate for a tidally locked planet: \begin{equation}
    S=\begin{cases}S_0\cos\vartheta\cos\left(\lambda-\lambda_0\right) & \text{on the day side}, \\ 
        0 & \text{on the night side}, 
    \end{cases}
\end{equation}
where $\lambda_0=0^{\circ}$  is the substellar longitude, and $S_0=1000\,\text{W\,m}^{-2}$. 

In the longwave band, upward and downward radiative fluxes are computed according to \begin{align}
    \frac{\text{d}F^{\uparrow}}{\text{d}\tau}&=F^{\uparrow}-\sigma T^{4} \\ 
    \frac{\text{d}F^{\downarrow}}{\text{d}\tau}&=\sigma T^{4}-F^{\downarrow}
\end{align} 
where $\tau=\tau_0(p/p_0)$ is the longwave optical depth, and we set $\tau_0=1$ and $p_0=p_{\text{s}}$. The longwave radiative heating in the model is then \begin{equation}
    \frac{\partial T}{\partial t} = \dots + \frac{g}{c_p}\frac{\partial\left(F^{\uparrow}-F^{\downarrow}\right)}{\partial p}.
\end{equation}
The boundary conditions for the longwave fluxes are $F^{\downarrow}(p=0)=0$ and $F^{\uparrow}(p=p_{\text{s}})=\sigma T_{\text{s}}^{4}$ where $T_{\text{s}}$ is the surface temperature.

\subsection{Surface energy budget, boundary layer sensible heat transport, and convection}

The surface is modeled as a static slab ocean that can exchange energy with the atmosphere in the vertical but cannot transport heat horizontally. The surface temperature evolves according to \begin{equation}
    C\frac{\partial T_{\text{s}}}{\partial t} = S + F^{\downarrow}_{\text{sfc}} - \sigma T^{4}_{\text{s}} - \mathcal{H} 
\end{equation}
where $C$ is a specified heat capacity that corresponds to an ocean mixed-layer depth of $2.5\,\text{m}$, $F^{\downarrow}_{\text{sfc}}$ is the downward flux of longwave radiation incident on the surface, and $\mathcal{H}$ is the surface sensible heat flux. $\mathcal{H}$ is computed according to \begin{equation}
    \mathcal{H}=\rho_{\text{a}}c_p\mathcal{C}\left\lvert\boldsymbol{u}_{\text{a}}\right\rvert\left(T_{\text{s}}-T_{\text{a}}\right),
\end{equation}
where $\rho$ is the density, the subscript `a' indicates quantities that are evaluated at the lowest model level, and $\mathcal{C}=0.001$ is a dimensionless bulk transfer coefficient. 

Turbulent vertical transport of heat within the boundary layer is parametrised following \citet{2016GMD.....9.1263T} as \begin{equation}
    \mathcal{D}_{T}=\overline{w^{\prime}\theta^{\prime}} = -\mathcal{K}\frac{\partial\theta}{\partial z},
\end{equation}
where $z$ is height, $w=\text{D}z/\text{D}t$ is the vertical velocity, and $\theta$ is the potential temperature. The diffusion coefficient $\mathcal{K}$ is calculated according to \begin{equation}
    \mathcal{K}=\begin{cases}
        \mathcal{C}\left\lvert\boldsymbol{u}_{\text{a}}\right\rvert z_{\text{a}} & \text{for}\ p>p_{\text{pbl}}, \\ 
        \mathcal{C}\left\lvert\boldsymbol{u}_{\text{a}}\right\rvert z_{\text{a}}\exp\left[-\left(\frac{p_{\text{pbl}-p}}{p_{\text{strat}}}\right)^{2}\right] & \text{for}\ p\leq p_{\text{pbl}}.
    \end{cases} 
\end{equation}
$p_{\text{pbl}}=850\,\text{hPa}$ is the top of the boundary layer, and $p_{\text{strat}}=100\,\text{hPa}$ controls the rate of decrease of boundary layer diffusion with height. The contribution to the model temperature tendency from the boundary layer diffusion is then \begin{equation}
    \frac{\partial T}{\partial t} = \cdots + \frac{g}{c_p}\frac{\partial\mathcal{D}_{T}}{\partial p}. 
\end{equation}

Finally, the model includes a dry convection scheme, which instantaneously restores the atmospheric temperature structure to the dry adiabat whenever it is convectively unstable.

\subsection{Surface friction and sponge layer}

Surface friction is parametrised following \citet{1994BAMS...75.1825H} as a linear Rayleigh drag \begin{equation}
    \frac{\partial\boldsymbol{u}}{\partial t} = \cdots - k_{\text{f}}\boldsymbol{u}
\end{equation}
where $k_{\text{f}}$ is defined as \begin{equation} 
    k_{\text{f}}=k_{\text{f,s}}\max\left(0,\frac{\sigma-\sigma_{\text{b}}}{1-\sigma_{\text{b}}}\right).
\end{equation}
$k_{\text{f,s}}=1\,\text{day}^{-1}$ is the drag timescale at the surface, and the top of the frictional boundary layer is located at $\sigma_{\text{b}}=0.7$. 

A linear drag is also applied in the upper atmosphere above $p_{\text{sponge}}=50\,\text{hPa}$ following \citet{2002GeoRL..29.1114P}, which damps the horizontal winds on a timescale of $0.5\ \text{days}^{-1}$. This acts as a sponge layer, and is included to suppress wave reflection at the model top.

\section{Derivation of expanded divergence equation}\label{sec:derive}

In this appendix we describe how the divergence equation (Equation \ref{eq:div1} in the main text) \begin{align}
    \frac{\partial\delta}{\partial t} + \nablap\cdot\left(\omega\frac{\partial\boldsymbol{u}}{\partial p}\right)+\nablap\cdot&\left[\left(f+\zeta\right)\mathbf{k}\times\boldsymbol{u}\right] \\ &=-\nablap^{2}\left(\phi+\frac{\left\lvert\boldsymbol{u}\right\rvert^{2}}{2}\right) -k_{\text{f}}\delta. \nonumber 
\end{align}
can be expanded into a form that separates contributions from the rotational and divergent winds (i.e., Equation \ref{eq:div_expand} in the main text). 

First, applying the Helmholtz decomposition to $\nablap\cdot[(f+\zeta)\mathbf{k}\times\boldsymbol{u}]$, we obtain four terms: \begin{enumerate}[leftmargin=.4cm]
    \setlength\itemsep{-1.13em}
    \item $\nablap\cdot[f\mathbf{k}\times\boldsymbol{u}_{\text{r}}] = \nablap\cdot[f\mathbf{k}\times\mathbf{k}\times\nablap\psi]=-\nablap\cdot[f\nablap\psi]$, \\
    \item $\nablap\cdot[\zeta\mathbf{k}\times\boldsymbol{u}_{\text{r}}] = -\nablap\cdot[\nablap^{2}\psi\nablap\psi]$, \\
    \item $\nablap\cdot[f\mathbf{k}\times\boldsymbol{u}_{\text{d}}]=\mathbf{k}\times\boldsymbol{u}_{\text{d}}\cdot\nablap f=u_{\text{d}}\beta$, \\ 
    \item $\nablap\cdot[\zeta\mathbf{k}\times\boldsymbol{u}_{\text{d}}]=\mathbf{k}\times\nablap\chi\cdot\nablap\zeta=\mathbf{k}\cdot(\nablap\chi\times\nablap\zeta)$\\[3pt]\phantom{.}\hfill$=J(\chi,\zeta)=J(\chi,\nablap^{2}\psi)$, 
\end{enumerate}
where $\beta=\partial f/\partial\vartheta$, for the third and fourth terms we have used $(f+\zeta)\nablap\cdot(\mathbf{k}\times\boldsymbol{u}_{\text{d}})=-(f+\zeta)(\nablap\times\boldsymbol{u}_{\text{d}})=0$, and in the fourth term we have defined the Jacobian \begin{equation}
    J(\chi,\zeta)\equiv\mathbf{k}\cdot\left(\nablap\chi\times\nablap\zeta\right).
\end{equation} 
We can also expand the $\lvert\boldsymbol{u}\rvert^{2}$ term into contributions from $\boldsymbol{u}_{\text{r}}$ and $\boldsymbol{u}_{\text{d}}$ as follows: \begin{equation}
    \frac{\left\lvert\boldsymbol{u}\right\rvert^{2}}{2}=\frac{\left\lvert\boldsymbol{u}_{\text{r}}\right\rvert^{2}}{2}+\frac{\left\lvert\boldsymbol{u}_{\text{d}}\right\rvert^{2}}{2}+\boldsymbol{u}_{\text{r}}\cdot\boldsymbol{u}_{\text{d}} = \frac{\left\lvert\nablap\psi\right\rvert^{2}}{2}+\frac{\left\lvert\nablap\chi\right\rvert^{2}}{2} + J\left(\psi,\chi\right).
\end{equation}

With these substitutions, the divergence Equation \eqref{eq:div1} becomes
\begin{widetext}
\begin{equation}
    \frac{\partial\delta}{\partial t}+\nablap\cdot\left[\omega\frac{\partial\left(\boldsymbol{u}_{\text{r}}+\boldsymbol{u}_{\text{d}}\right)}{\partial p}\right]+u_{\text{d}}\beta+J\left(\chi,\nablap^{2}\psi\right)-\nablap\cdot\left[\left(f+\nablap^{2}\psi\right)\nablap\psi\right]=-\nablap^{2}\left[\phi+\frac{\left\lvert\nablap\psi\right\rvert^{2}}{2}+\frac{\left\lvert\nablap\chi\right\rvert^{2}}{2}+J\left(\psi,\chi\right)\right] - k_{\text{f}}\delta, \label{eq:div_expand2}
\end{equation}
\end{widetext}
which is Equation \eqref{eq:div_expand} in the main text.

\clearpage 

\section{Subdivision of height field}\label{sec:decompose}

This appendix includes some additional analysis of the terms in the divergence equation (Equation \ref{eq:div_expand}) for our simulations, to identify those that contribute most strongly to the rotational and divergent height components. In particular, we analyse the divergent circulation by highlighting the term that corresponds to overturning circulation, and comparing it to the terms that correspond to wave-like divergent circulation and non-linear rotational-divergent interactions

\begin{figure}[!b]
    \centering\includegraphics[width=.41\textwidth]{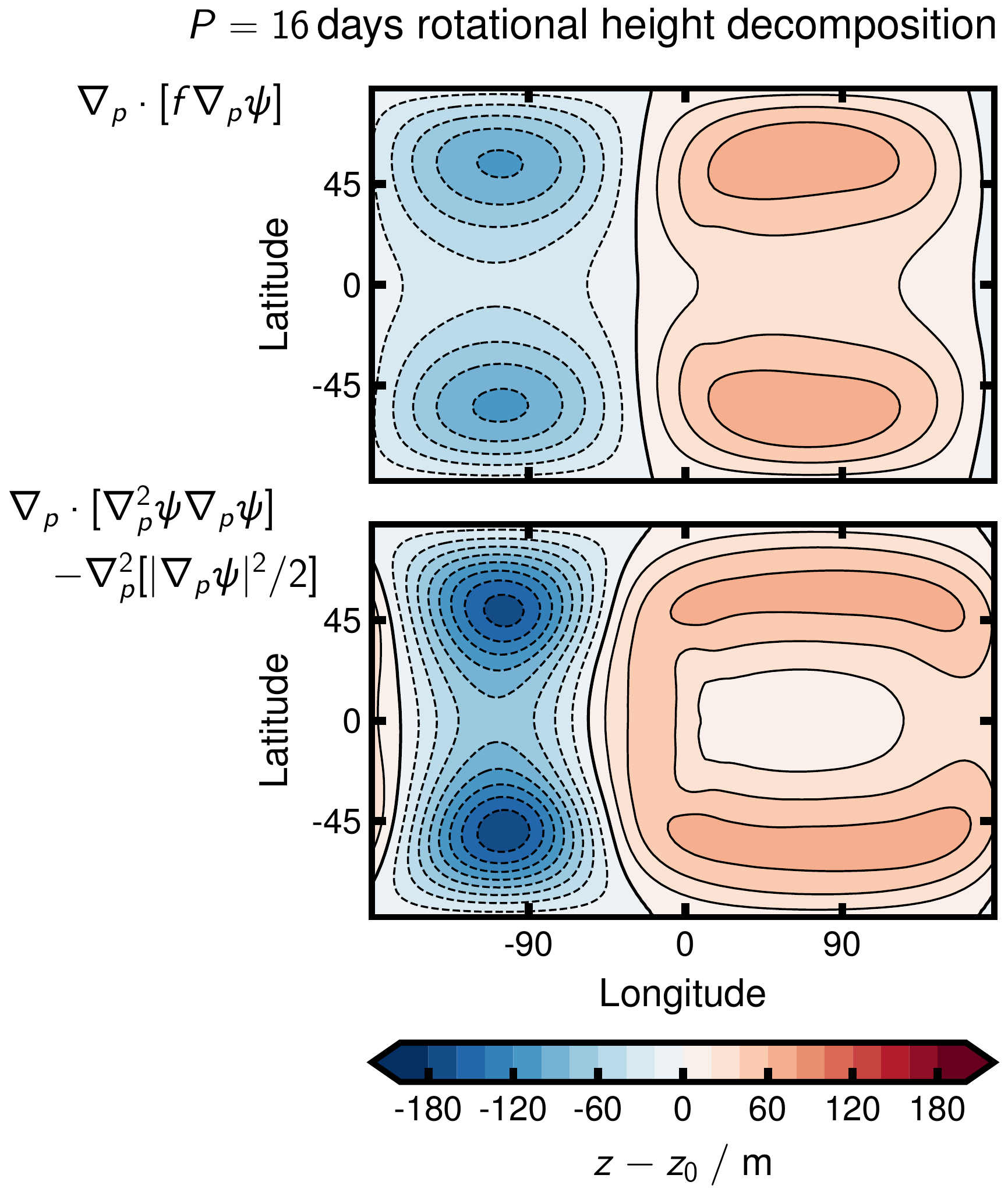}
    \caption{Subdivision of rotational geopotential height in the $P=16\ \text{days}$ simulation into contributions from linear (top panel) and non-linear (bottom panel) terms.}\label{fig:rot_decomp}
\end{figure}

In the $P=4$ and $P=8\ \text{days}$ simulations, the eddy rotational height (Figure \ref{fig:uz480}) is dominated by the linear $\nablap\cdot[f\nablap\psi]$ term (not shown). It has the structure of a stationary equatorial Rossby wave, that is Doppler-shifted eastwards by the zonal-mean jet \citep{2014ApJ...793..141T,2018ApJ...869...65H}. In the $P=16\ \text{days}$ simulation, the eddy rotational height also resembles a stationary Rossby wave, but the lobes with a positive geopotential anomaly are noticeably elongated with respect to those with a negative geopotential anomaly (Figure \ref{fig:uz480}). Figure \ref{fig:rot_decomp} show the eddy rotational height field for the $P=16\ \text{days}$ simulation, divided into contributions from the linear term, and the non-linear $\nablap\cdot[\nablap^{2}\nablap\psi]-\nablap^{2}[\lvert\nablap\psi\rvert^{2}/2]$ terms. This subdivision of the height field shows that the elongation of the stationary Rossby is mostly due to non-linear rotational-rotational interactions. 

\begin{figure}[!t]
    \centering\includegraphics[width=.41\textwidth]{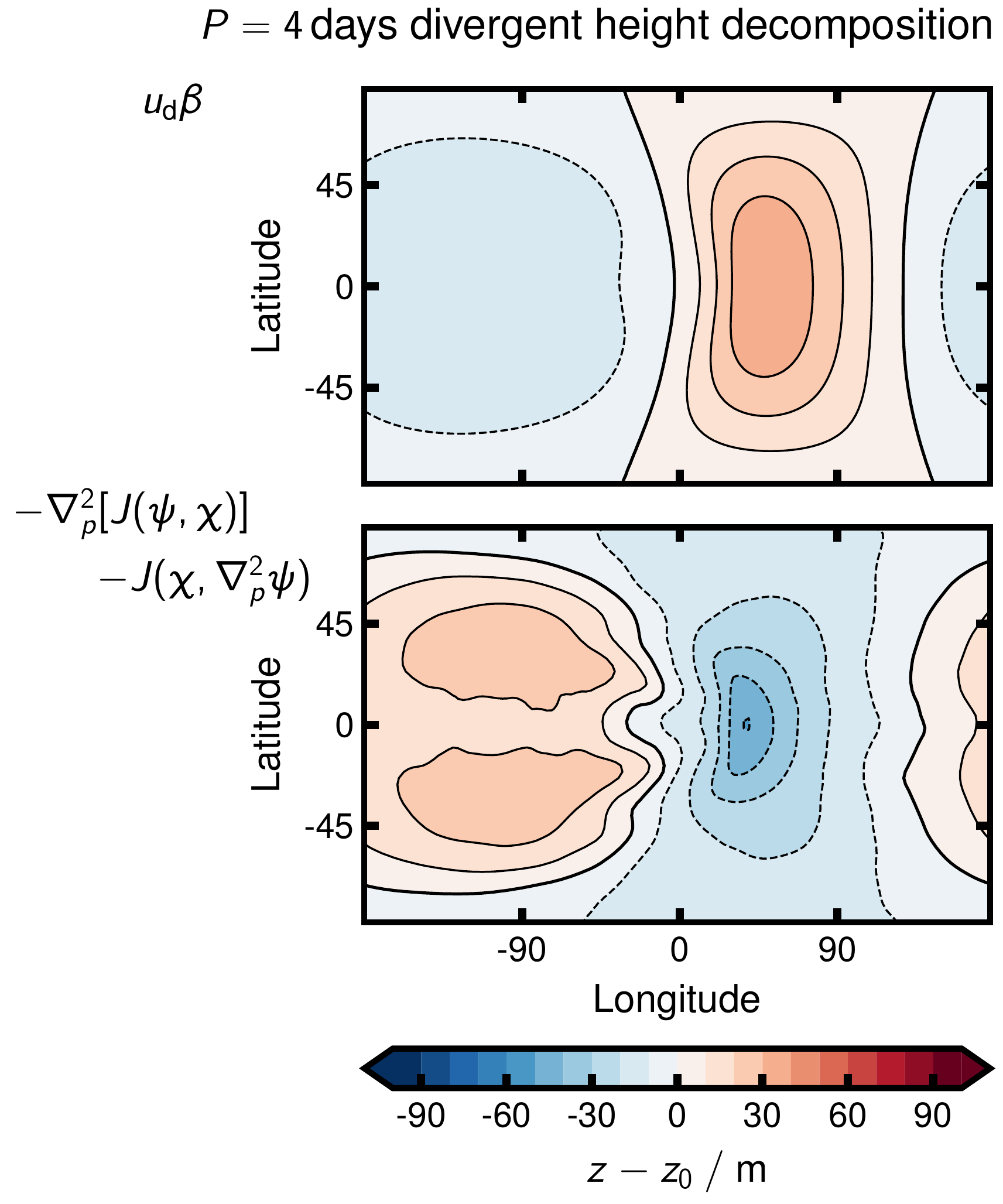}
    \caption{Subdivision of divergent geopotential height in the $P=4\ \text{days}$ simulation into contributions from the linear $u_{\text{d}}\beta$ term (top panel) and non-linear rotational-divergent terms (bottom panel).}\label{fig:div_decomp}
\end{figure}

\begin{figure*}
    \centering\includegraphics[width=.95\textwidth]{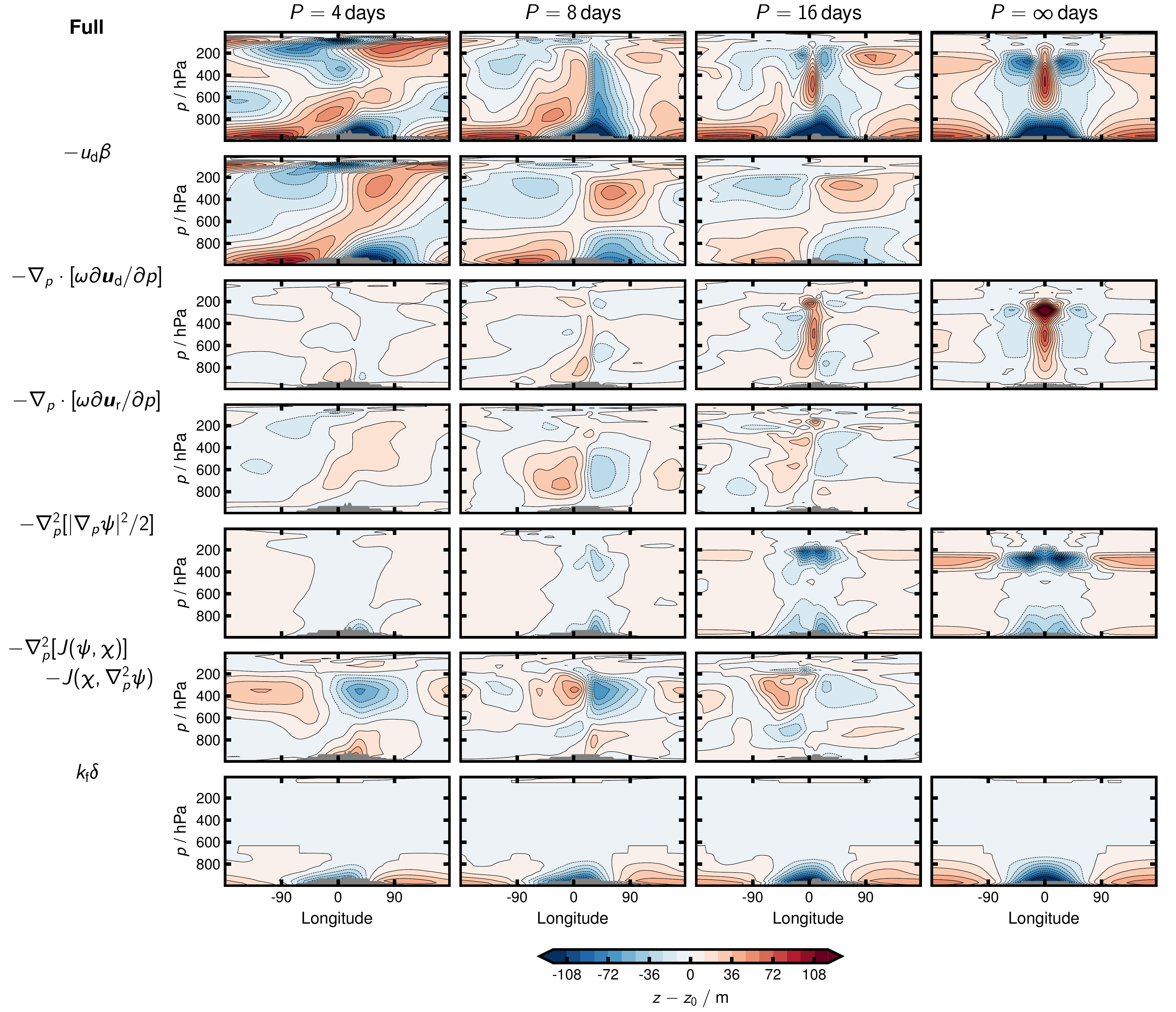}
    \caption{Subdivision of divergent geopotential height into contributions from different terms in the divergence equation (Equation \ref{eq:div_expand} in the main text). The top row shows the total divergent height field for each simulation, and subsequent rows show contributions to the divergent height from different terms (labelled to the left of each row). Contributions from terms in the full divergence equation that are due to solely rotational circulation are not shown, as they do not contribute to the divergent geopotential by definition.}\label{fig:div_vert_decomp}
\end{figure*}

The divergent height field has a complex structure in each of the simulations presented in Figures \ref{fig:uz480} and \ref{fig:uzlonp}. In order to interpret it, we show the divergent height field decomposed into contributing terms in Figures \ref{fig:div_decomp} and \ref{fig:div_vert_decomp}. 

The $P=\infty\ \text{days}$ simulation represents a limiting case where the divergent circulation takes the form of an isotropic, overturning cell (see Figures \ref{fig:uz480} and \ref{fig:uzlonp}). In this scenario, Figure \ref{fig:div_vert_decomp} shows that the divergent height structure is due to the $\nablap\cdot(\omega\partial\boldsymbol{u}_{\text{d}}\partial p)$ term in regions of ascent near the substellar point, and descent on the night side. Meanwhile, the $k_{\text{f}}\delta$ (surface friction) contribution is important in the boundary layer, and non-linear horizontal divergent-divergent interactions are important in the outflow area surrounding the region of ascent near the substellar point. The balance of terms in the $P=16\ \text{days}$ simulation is similar to that in the $P=\infty\ \text{days}$ case, although there is a contribution to the divergent height field from the horizontal non-linear rotational-divergent interaction terms, $-\nablap^{2}[J(\psi,\chi)]-J(\chi,\nablap^{2}\psi)$ (mostly through $\overline{u}_{\text{r}}$). Our interpretation of the divergent height structure in the $P=16\ \text{days}$ simulation is that it is consistent with a thermally-direct overturning circulation that is `tipped over' by the superrotating jet. 

By contrast, the most important contributions to the divergent height in the $P=4\ \text{days}$ simulation are from the linear $u_{\text{d}}\beta$ term, and non-linear horizontal rotational-divergent interactions (note that surface friction is still important in the boundary layer). Figure \ref{fig:div_decomp} shows the horizontal structure of the linear $u_{\text{d}}\beta$ term and horizontal non-linear rotational-divergent interaction terms for the $P=4\ \text{days}$ simulation. The height field associated with the $u_{\text{d}}\beta$ term is qualitatively similar to that of the linear Kelvin wave obtained in the linear shallow water simulation presented in Appendix \ref{sec:sw}. In the vertical, the height associated with this term has a wavenumber-1 structure (Figure \ref{fig:div_vert_decomp}). However, unlike in the linear shallow water simulation, the divergent height field is modified by the contribution from the non-linear rotational-divergent interaction term (largest in the upper troposphere), which yields the complex structure shown in Figure \ref{fig:uz480}. The divergent height in the $P=8\ \text{days}$ simulation has contributions from all of the terms discussed for both the $P=4\ \text{days}$ and $P=16\ \text{days}$ simulations, and it appears that this simulation is an intermediate case between the Kelvin/non-linear wave regime of the $P=4\ \text{days}$ simulation, and the thermally-direct overturning regime of the $P=16$ and $P=\infty\ \text{days}$ simulations. 

\newpage

\bibliography{bibliography}{}
\bibliographystyle{aasjournal}

\end{document}